# ARTICLE

# von Laue's theorem and its applications

Changbiao Wang

**Abstract:** von Laue's theorem, as well as its generalized form, is strictly proved in detail for its sufficient and necessary condition (SNC). This SNC version of Laue's theorem is used to analyze the infinitely extended electrostatic field produced by a charged metal sphere in free space, and the static field confined in a finite region of space. It is shown in general that the total (Abraham=Minkowski) electromagnetic momentum and energy for the electrostatic field cannot constitute a Lorentz four-vector. A derivative von Laue's theorem, which provides a criterion for a Lorentz invariant, is also presented.

**Résumé :** Nous démontrons en détail le théorème de von Laue et sa généralisation sur les conditions nécessaires et suffisantes. Nous utilisons cette version du théorème de von Laue pour analyser le champ électrique produit par une sphère métallique chargée dans l'espace infini et dans une région confinée de l'espace. Nous montrons que les composantes totales (Abraham=Minkowski) de moment et d'énergie EM ne constituent pas un quatre vecteur de Lorentz. Nous présentons aussi un dérivé du théorème de von Laue qui fournit le critère pour un invariant de Lorentz. [Traduit par la Rédaction]

## 1. Introduction

This paper is trying to solve one of the most controversial problems in classical physics, so-called von Laue's theorem [1]. This theorem is well known in the dynamics of relativity, and it is widely presented in textbooks and literature [2–10], but often with different explanations or understandings for its sufficient and necessary conditions so that several additional versions of Laue's theorem were born [2–4].

von Laue's theorem [1] provides a criterion to judge whether the space integrals of the time-row (column) elements of a tensor constitute a Lorentz *four-vector*. In some publications, the divergence-less property of a tensor (Landau–Lifshitz) [2], or the divergence-less plus a symmetry (Weinberg) [3], or the divergence-less plus an implicit zero-boundary condition (Møller) [4] is taken as a sufficient condition,[1,2,3,4] while in some others the divergence-less is taken as a necessary condition [7, 8].

According to the divergence property of the "electromagnetic (EM)" or Poincaré "complete" stress–energy tensor of the classical electron, Laue's theorem has been used for identifying whether its total momentum and energy constitute a four-vector [7].[5] In a recent Letter, the theorem is also implicitly applied to relativistic analysis of the dielectric Einstein-box thought experiment for resolution of the Abraham–Minkowski debate on light momentum in a medium [10].[6]

A theorem usually has two parts: a condition and a conclusion. A sufficient condition can be used to affirm the theorem's conclusion while it cannot be used to negate the conclusion. In contrast, a necessary condition can be used to negate the theorem's conclusion while it cannot be used to affirm the conclusion. Only a sufficient and necessary condition (SNC) can be used for both. However, to our best knowledge, the necessary condition for Laue's theorem has never been proved in the previous publications, including Laue's original work [1].

In this paper, a strict proof of the SNC version of Laue's theorem (Sect. 2), as well as its generalized form (Sect. 7), is given, and it is shown that the divergence-less itself is *neither* a sufficient *nor* a necessary condition, while the divergence-less plus an additional boundary condition *only* can be a sufficient condition (Sect. 6).

---

**C. Wang.** ShangGang Group, 70 Huntington Road, Apartment 11, New Haven, CT 06512, USA.
**E-mail for correspondence:** changbiao_wang@yahoo.com.

[1]In ref. 2, the book by Landau and Lifshitz, the *divergence-less* of a tensor is taken as a sufficient condition, as shown in Eq. (32.6) on p. 83 and Eq. (32.11) on p. 84. The *symmetry* of the tensor is claimed to be required by "the law of conservation of angular momentum" by repeating use of their version of Laue's theorem; see Eq. (32.10) on p. 84. As shown in Sect. 4 of the present paper, however, the divergence-less is *never* a sufficient condition; thus the correctness of the requirement of the symmetry is also questionable.

[2]On p. 46 of ref. 3, Weinberg argues that it "can be shown" that the divergence-less and symmetry is a sufficient condition.

[3]On pp. 166–169 of ref. 4, Møller provided a proof that the divergence-less property plus an implicit "zero-boundary condition" is a sufficient condition. This zero-boundary condition, combined with the divergence-less, insures that the space integrals of the time-column elements "are constant in time", as shown in Eq. (24) on p. 167. The argument of the zero-boundary condition is "the system considered is finite"; however, in practice, a finite system does not necessarily mean a zero-boundary condition, of which a typical example is given in Fig. 1 of the present paper. Thus the use of Møller's version of Laue's theorem is very limited; for example, it cannot be used to judge the Lorentz property of EM momentum and energy of the charged metal sphere; see Sect. 4 of the present paper.

[4]On p. 756 of ref. 5, Jackson presented Landau–Lifshitz version of Laue's theorem [2], where the divergence-less described by Eq. (16.39) is taken as a sufficient condition, and it is thought to be equivalent to the original Laue's sufficient condition Eq. (16.40). In fact, Eq. (16.39) does not necessarily mean Eq. (16.40), as shown in Sect. 4 of the present paper.

[5]In ref. 7, the divergence-less is claimed as a sufficient condition in Eq. (19), while it is taken as the necessary condition in Eq. (51).

[6]In ref. 10, as shown in Eq. (5), the symmetry and divergence-less of a four-tensor is implicitly taken as a sufficient condition for the space integrals of the time-column elements to constitute a Lorentz four-vector; namely, the Weinberg's version of Laue's theorem.

As an application, the SNC version of Laue's theorem is used to analyze the infinitely extended electrostatic field produced by a charged metal sphere in free space (Sect. 4), and the electrostatic field that is confined in a finite region of space (Sect. 5). Finally, a derivative von Laue's theorem (Sect. 8), which provides a criterion for a Lorentz *invariant*, is also presented as well.

## 2. von Laue's theorem for a Lorentz four-vector

**von Laue's theorem.** Assume that $\Theta^{\mu\nu}(\mathbf{x})$ is a Lorentz four-tensor given in the laboratory frame $XYZ$ ($\mu, \nu = 1, 2, 3,$ and $4$, with the index 4 corresponding to time component), $\Theta^{\mu\nu}$ is independent of time ($\partial \Theta^{\mu\nu}/\partial t \equiv 0$), and further, $\Theta^{i4}(\mathbf{x}) = 0$ holds for all $i = 1, 2,$ and $3$. von Laue's theorem [1] states: *The time-row-element space integrals (which are assumed to be convergent)*

$$P^\nu = \int_V \Theta^{4\nu} d^3x \tag{1}$$

*constitute a Lorentz four-vector if and only if*

$$\int_V \Theta^{ij} d^3x = 0 \tag{2}$$

*holds for all $i, j = 1, 2,$ and $3$.*[7]

Equation (2) is a sufficient and necessary condition of Laue's theorem. For clarity, the theorem is illustrated as follows:

$$\Theta^{\mu\nu}(\mathbf{x}) = \begin{pmatrix} \Theta^{ij} & \Theta^{i4} = 0 \\ \Theta^{4j} & \Theta^{44} \end{pmatrix}: \quad \int_V \Theta^{ij} d^3x = 0$$

$$\Leftrightarrow \quad P^\nu = \int_V \Theta^{4\nu} d^3x \quad \text{to be a four-vector} \tag{3}$$

Note the positions of the superscript index 4 in $\Theta^{i4} = 0$ and $P^\nu = \int_V \Theta^{4\nu} d^3x$, as shown in (3). $\Theta^{\mu\nu}(\mathbf{x})$ is not necessarily symmetric. The integral domain $V$ can be multiply connected.

As shown later in this paper, (2) and $\partial_\mu \Theta^{\mu\nu}(\mathbf{x}) = 0$ (divergence-less) are *not* equivalent; namely, $\int_V \Theta^{ij}(\mathbf{x}) d^3x = 0$ does not derive $\partial_\mu \Theta^{\mu\nu}(\mathbf{x}) = 0$, and $\partial_\mu \Theta^{\mu\nu}(\mathbf{x}) = 0$ does not derive $\int_V \Theta^{ij}(\mathbf{x}) d^3x = 0$. However, $\partial_\mu \Theta^{\mu\nu}(\mathbf{x}) = 0$ (divergence-less) plus an additional boundary condition can be a sufficient condition.

In (1), we designate the time-*row*-element space integrals $P^\nu = \int_V \Theta^{4\nu} d^3x$ as a four-vector. Alternatively, we also can designate the time-*column*-element space integrals $\Pi^\mu = \int_V \Theta^{\mu 4} d^3x$ as a four-vector, for which the statement is modified into: Assume that $\Theta^{\mu\nu}(\mathbf{x})$ is a Lorentz four-tensor given in the laboratory frame $XYZ$ ($\mu, \nu = 1, 2, 3,$ and $4$, with the index 4 corresponding to time component), $\Theta^{\mu\nu}$ is independent of time ($\partial \Theta^{\mu\nu}/\partial t \equiv 0$), and further, $\Theta^{4j}(\mathbf{x}) = 0$ holds for all $j = 1, 2,$ and $3$. von Laue's theorem states: *The time-column-element space integrals $\Pi^\mu = \int_V \Theta^{\mu 4} d^3x$ constitute a Lorentz four-vector if and only if $\int_V \Theta^{ij} d^3x = 0$ holds for all $i, j = 1, 2,$ and $3$*. Obviously, if $\Theta^{\mu\nu}(\mathbf{x})$ is symmetric, $P^\nu$ and $\Pi^\mu$ are equal. Without loss of generality, only a proof for the row-four-vector case is given here.

*Proof*: Suppose that $X'Y'Z'$ is an inertial frame moving at $\boldsymbol{\beta}c$ with respect to the laboratory frame $XYZ$, where $c$ is the vacuum light speed. The time–space Lorentz transformation is given by [5]

$$\mathbf{x}' = \mathbf{x} + \xi(\boldsymbol{\beta} \cdot \mathbf{x})\boldsymbol{\beta} - \gamma \boldsymbol{\beta} ct \tag{4}$$

$$ct' = \gamma(ct - \boldsymbol{\beta} \cdot \mathbf{x}) \tag{5}$$

where $\xi \equiv (\gamma - 1)/\boldsymbol{\beta}^2$ and $\gamma \equiv (1 - \boldsymbol{\beta}^2)^{-1/2}$.

According to (1), we define $P'^\nu = \int \Theta'^{4\nu}(\mathbf{x}', ct') d^3x'$ in $X'Y'Z'$, and we will show that $P'^\nu$ and $P^\nu$ follow the four-vector Lorentz transformation if and only if $\int \Theta^{ij} d^3x = 0$ holds.

With the time–space four-vector given by $X^\mu \equiv (\mathbf{x}, ct)$, from the Lorentz transformation of $\Theta'^{4\nu}(\mathbf{x}', ct')$ we have

$$P'^\nu = \int_{V': t'=\text{const}} \Theta'^{4\nu}(\mathbf{x}', ct') d^3x'$$

$$= \frac{\partial X'^4}{\partial X^\sigma} \frac{\partial X'^\nu}{\partial X^\lambda} \int_{V': t'=\text{const}} \Theta^{\sigma\lambda}(\mathbf{x} = \mathbf{x}(\mathbf{x}', ct')) d^3x'$$

$$= \frac{\partial X'^4}{\partial X^\sigma} \frac{\partial X'^\nu}{\partial X^\lambda} \gamma^{-1} \int_V \Theta^{\sigma\lambda}(\mathbf{x}) d^3x$$

$$\left(\text{Note: } \frac{\partial P'^\nu}{\partial t'} \equiv 0 \quad \text{because} \quad \frac{\partial}{\partial X^\mu} \int_V \Theta^{\sigma\lambda}(\mathbf{x}) d^3x \equiv 0\right)$$

$$= \frac{\partial X'^\nu}{\partial X^\lambda} \left[\frac{\partial X'^4}{\partial X^4} \gamma^{-1} \int_V \Theta^{4\lambda}(\mathbf{x}) d^3x + \frac{\partial X'^4}{\partial X^i} \gamma^{-1} \int_V \Theta^{i\lambda}(\mathbf{x}) d^3x\right]$$

$$\text{(with } i = 1, 2, 3\text{)}$$

$$= \frac{\partial X'^\nu}{\partial X^\lambda} \int_V \Theta^{4\lambda}(\mathbf{x}) d^3x + \frac{\partial X'^\nu}{\partial X^\lambda} \frac{\partial X'^4}{\partial X^i} \gamma^{-1} \int_V \Theta^{i\lambda}(\mathbf{x}) d^3x$$

$$\left(\text{because } \frac{\partial X'^4}{\partial X^4} \gamma^{-1} = \frac{\partial ct'}{\partial ct} \gamma^{-1} = 1\right)$$

$$= \frac{\partial X'^\nu}{\partial X^\lambda} P^\lambda + \frac{\partial X'^\nu}{\partial X^\lambda} \frac{\partial X'^4}{\partial X^i} \gamma^{-1} \int_V \Theta^{i\lambda}(\mathbf{x}) d^3x$$

$$\left(\text{from definition, } P^\lambda = \int_V \Theta^{4\lambda}(\mathbf{x}) d^3x, \text{ with } \frac{\partial P^\lambda}{\partial t} \equiv 0\right)$$

$$= \frac{\partial X'^\nu}{\partial X^\lambda} P^\lambda + \frac{\partial X'^\nu}{\partial X^4} \frac{\partial X'^4}{\partial X^i} \gamma^{-1} \int_V \Theta^{i4}(\mathbf{x}) d^3x + \frac{\partial X'^\nu}{\partial X^j} \frac{\partial X'^4}{\partial X^i} \gamma^{-1} \int_V \Theta^{ij}(\mathbf{x}) d^3x \tag{6}$$

$$\text{(with } i, j = 1, 2, 3\text{)}$$

$$= \frac{\partial X'^\nu}{\partial X^\lambda} P^\lambda + \frac{\partial X'^\nu}{\partial X^j} \frac{\partial X'^4}{\partial X^i} \gamma^{-1} \int_V \Theta^{ij}(\mathbf{x}) d^3x$$

$$\text{(because } \Theta^{i4}(\mathbf{x}) = 0 \text{ for } i = 1, 2, 3\text{)}$$

$$= \frac{\partial X'^\nu}{\partial X^\lambda} P^\lambda + \frac{\partial X'^\nu}{\partial X^i} \frac{\partial X'^4}{\partial X^j} \gamma^{-1} \int_V \Theta^{ji}(\mathbf{x}) d^3x$$

$$\text{(exchanging the dummy indices } j \text{ and } i\text{)}$$

From (6), we know that the sufficient and necessary conditions for the validity of $P'^\nu = (\partial X'^\nu/\partial X^\lambda) P^\lambda$ is the holding of

$$\gamma^{-1} \frac{\partial X'^\nu}{\partial X^i} \left[\int_V \Theta^{ji}(\mathbf{x}) d^3x\right] \frac{\partial X'^4}{\partial X^j} = 0 \tag{7}$$

or

---

[7] In the original ref. 1, Laue only gave the proof that $\int \Theta^{ij}(\mathbf{x}) d^3x = 0$ is a *sufficient* condition. In principle, this original Laue's theorem cannot be used to judge the Lorentz property of EM momentum and energy of the charged metal sphere, because the EM stress–energy tensor for the charged metal sphere does not meet $\int \Theta^{ij}(\mathbf{x}) d^3x = 0$. Thus one of the significant contributions of the present paper is to show that $\int \Theta^{ij}(\mathbf{x}) d^3x = 0$ is also a *necessary* condition, and consequently, the difficulty that the original Laue's theorem has is resolved; confer Sect. 4.

$$\begin{pmatrix} 1+\xi\beta_x^2 & \xi\beta_x\beta_y & \xi\beta_x\beta_z \\ \xi\beta_y\beta_x & 1+\xi\beta_y^2 & \xi\beta_y\beta_z \\ \xi\beta_z\beta_x & \xi\beta_z\beta_y & 1+\xi\beta_z^2 \\ -\gamma\beta_x & -\gamma\beta_y & -\gamma\beta_z \end{pmatrix}$$

$$\times \begin{pmatrix} a_{11} & a_{12} & a_{13} \\ a_{21} & a_{22} & a_{23} \\ a_{31} & a_{32} & a_{33} \end{pmatrix} \begin{pmatrix} -\gamma\beta_x \\ -\gamma\beta_y \\ -\gamma\beta_z \end{pmatrix} = \begin{pmatrix} 0 \\ 0 \\ 0 \\ 0 \end{pmatrix} \quad (8)$$

where $a_{ij} = \int_V \Theta^{ji}(\mathbf{x}) d^3x$, with altogether nine elements but only four independent linear equations. However $\boldsymbol{\beta}c$ is arbitrary, and $a_{ij} = 0$ must hold for all $i, j = 1, 2$, and 3. For example, $(\beta_x \neq 0, \beta_y = 0, \beta_z = 0) \Rightarrow a_{i1} = 0$, $(\beta_x = 0, \beta_y \neq 0, \beta_z = 0) \Rightarrow a_{i2} = 0$, and $(\beta_x = 0, \beta_y = 0, \beta_z \neq 0) \Rightarrow a_{i3} = 0$, all for $i = 1, 2$, and 3. Inversely if $a_{ij} = 0$ holds, then (7) must hold, leading to the validity of $P'^\nu = (\partial X'^\nu/\partial X^\lambda)P^\lambda$ from (6). Thus we finish the proof of the necessity and sufficiency.

From (6), we can see that the sufficiency of $\int \Theta^{ij} d^3x = 0$ is apparent; however, to show its necessity, we must employ the attribute of four-vector: If $P^\nu$ is a four-vector, then it follows Lorentz transformation between *any two* of inertial frames (namely, $\boldsymbol{\beta}c$ is arbitrary).

The importance of pre-assumption $\partial\Theta^{\mu\nu}/\partial t \equiv 0$ in $XYZ$ frame should be emphasized, which ensures that the integral $P'^\nu = \int \Theta'^{4\nu}(\mathbf{x}', ct') d^3x'$ in $X'Y'Z'$ frame does not depend on $t'$.

It is worthwhile to point out that, the sufficient and necessary conditions given by Jammer [6] and Bialynicki-Birula [9] are not equivalent to (2).[8,9]

## 3. EM stress–energy tensor

From the definition $F^{\mu\nu} = \partial^\mu A^\nu - \partial^\nu A^\mu$ and the Maxwell equation $\partial_\mu G^{\mu\nu} = J^\nu$ resulting from $[\nabla \times \mathbf{H} - \partial(c\mathbf{D})/\partial(ct), \nabla \cdot (c\mathbf{D})] = (\mathbf{J}, c\rho)$, we obtain

$$F^{\mu\nu} = \begin{pmatrix} 0 & -B_z & B_y & E_x/c \\ B_z & 0 & -B_x & E_y/c \\ -B_y & B_x & 0 & E_z/c \\ -E_x/c & -E_y/c & -E_z/c & 0 \end{pmatrix} \quad (9)$$

$$G_{\mu\nu} = g_{\mu\sigma}G^{\sigma\lambda}g_{\lambda\nu} = \begin{pmatrix} 0 & -H_z & H_y & -D_xc \\ H_z & 0 & -H_x & -D_yc \\ -H_y & H_x & 0 & -D_zc \\ D_xc & D_yc & D_zc & 0 \end{pmatrix} \quad (10)$$

where $g^{\mu\nu} = g_{\mu\nu} = \text{diag}(-1, -1, -1, +1)$ is the Minkowski metric. The EM stress–energy tensor is defined as

$$T^{\mu\nu} = g^{\mu\sigma}G_{\sigma\lambda}F^{\lambda\nu} + \frac{1}{4}g^{\mu\nu}G_{\sigma\lambda}F^{\sigma\lambda} \quad \text{or} \quad T^{\mu\nu} = \begin{pmatrix} \check{\mathbf{T}}_M & c\mathbf{g}_A \\ c\mathbf{g}_M & W_{em} \end{pmatrix} \quad (11)$$

where $\mathbf{g}_A = \mathbf{E} \times \mathbf{H}/c^2$ is the Abraham momentum, $\mathbf{g}_M = \mathbf{D} \times \mathbf{B}$ is the Minkowski momentum, $W_{em} = 0.5(\mathbf{D} \cdot \mathbf{E} + \mathbf{B} \cdot \mathbf{H})$ is the EM energy density, and $\check{\mathbf{T}}_M = -\mathbf{D}\mathbf{E} - \mathbf{B}\mathbf{H} + \check{\mathbf{I}}0.5(\mathbf{D} \cdot \mathbf{E} + \mathbf{B} \cdot \mathbf{H})$ is the Minkowski stress tensor, with $\check{\mathbf{I}}$ the unit tensor. $G_{\sigma\lambda}F^{\sigma\lambda} = 2(\mathbf{B} \cdot \mathbf{H} - \mathbf{D} \cdot \mathbf{E})$ is Lorentz invariant.

If the EM field has the following properties (in a uniform isotropic medium, for example)

$$\mathbf{D} \times (\nabla \times \mathbf{E}) = \mathbf{E} \times (\nabla \times \mathbf{D}) \qquad \mathbf{D} \cdot \nabla \mathbf{E} = \mathbf{E} \cdot \nabla \mathbf{D} \quad (12)$$

$$\mathbf{B} \times (\nabla \times \mathbf{H}) = \mathbf{H} \times (\nabla \times \mathbf{B}) \qquad \mathbf{B} \cdot \nabla \mathbf{H} = \mathbf{H} \cdot \nabla \mathbf{B} \quad (13)$$

$$\frac{\partial \mathbf{B}}{\partial t} \cdot \mathbf{H} = \frac{\partial \mathbf{H}}{\partial t} \cdot \mathbf{B} \qquad \frac{\partial \mathbf{E}}{\partial t} \cdot \mathbf{D} = \frac{\partial \mathbf{D}}{\partial t} \cdot \mathbf{E} \quad (14)$$

leading to

$$G_\lambda^\mu \partial_\mu F^{\lambda\nu} + \frac{1}{4}\partial^\nu(G_{\sigma\lambda}F^{\sigma\lambda}) = 0 \quad (15)$$

then we obtain the well-known forms of momentum and energy conservation equations, given by

$$\partial_\mu T^{\mu\nu} = F^{\mu\nu}J_\mu + \left[G_\lambda^\mu \partial_\mu F^{\lambda\nu} + \frac{1}{4}\partial^\nu(G_{\sigma\lambda}F^{\sigma\lambda})\right] = F^{\mu\nu}J_\mu \quad (16)$$

or

$$\nabla \cdot \check{\mathbf{T}}_M + \frac{\partial(\mathbf{D} \times \mathbf{B})}{\partial t} = -\rho\mathbf{E} - \mathbf{J} \times \mathbf{B} \quad (17)$$

and

$$\nabla \cdot (\mathbf{E} \times \mathbf{H}) + \frac{\partial W_{em}}{\partial t} = -\mathbf{E} \cdot \mathbf{J} \quad (18)$$

Equations (15) and (16) are pure tensor equations. Accordingly, if (15) and (16) hold in one inertial frame, they hold in all inertial frames. For a uniform isotropic medium observed in the medium-rest frame, (15) holds and we have $\partial_\mu T^{\mu\nu} = F^{\mu\nu}J_\mu$ holding; thus leading to the holding of $\partial_\mu T^{\mu\nu} = F^{\mu\nu}J_\mu$ in all inertial frames.

## 4. Application of Laue's theorem to charged metal sphere in free space

Consider an infinitely extended static field in free space, produced by a charged metal sphere in free space. It is assumed that the metal sphere is made a perfect conductor, and thus the basic electrostatic properties of ideal conductors apply [11]. In free space, $\mathbf{D} = \varepsilon_0\mathbf{E}$ and $\mathbf{B} = \mu_0\mathbf{H}$ hold, where $\varepsilon_0$ and $\mu_0$ are the vacuum permittivity and permeability, respectively. In such a case, $\check{\mathbf{T}}_M = -\mathbf{D}\mathbf{E} - \mathbf{B}\mathbf{H} + \check{\mathbf{I}}0.5(\mathbf{D} \cdot \mathbf{E} + \mathbf{B} \cdot \mathbf{H})$ is symmetric, and $\mathbf{g}_A = \mathbf{g}_M$ or $\mathbf{E} \times \mathbf{H}/c^2 = \mathbf{D} \times \mathbf{B}$ holds, namely, the Abraham and Minkowski EM momentum densities are equal. Thus from (11) we know that the EM stress–energy tensor $T^{\mu\nu}$ for the charged metal sphere is *symmetric*.

$\partial T^{\mu\nu}/\partial t \equiv 0$ and $T^{i4} = 0$ ($c\mathbf{g}_A = 0$) hold in the metal sphere rest frame, but from (11) we have $T^{ii} = T^{11} + T^{22} + T^{33} = 0.5\varepsilon_0\mathbf{E}^2$ leading to $\int T^{ii} d^3x \neq 0$ so that $\int T^{ij} d^3x = 0$ cannot hold for all $i, j = 1, 2$, and 3. Thus according to Laue's theorem, we judge that the time-row-element space integrals

---

[8]On p. 197 of ref. 6 by Jammer, the vanishing of all of the integrated *diagonal* elements of $\Theta^{ij}(\mathbf{x})$, namely, $\int_V \Theta^{11} d^3x = \int_V \Theta^{22} d^3x = \int_V \Theta^{33} d^3x = 0$, is taken as a sufficient and necessary condition for Laue's theorem. Obviously, $\int \Theta^{ij} d^3x = 0$ derives $\int_V \Theta^{11} d^3x = \int_V \Theta^{22} d^3x = \int_V \Theta^{33} d^3x = 0$, but the latter does not derive the former.

[9]In ref. 9 by Bialynicki-Birula, the vanishing of the integrated trace of $\Theta^{ij}(\mathbf{x})$, namely, $\int_V \Theta^{ii} d^3x = \int_V (\Theta^{11} + \Theta^{22} + \Theta^{33}) d^3x = 0$, is taken as a sufficient and necessary condition for Laue's theorem. Obviously, $\int \Theta^{ij} d^3x = 0$ derives $\int_V \Theta^{ii} d^3x = 0$, but $\int_V \Theta^{ii} d^3x = 0$ does not derive $\int \Theta^{ij} d^3x = 0$.

$$\int T^{4\nu}d^3x = \left(\int c\mathbf{g}_M d^3x, \int W_{em}d^3x\right)$$

or

$$\left[\int (\mathbf{D}\times\mathbf{B})d^3x,\ \frac{1}{c}\int W_{em}d^3x\right] \quad (19)$$

cannot constitute a Lorentz four-vector.

It is interesting to point out that the versions of Laue's theorem, where $\partial_\mu\Theta^{\mu\nu}=0$ (Landau–Lifshitz version) [2] or $\partial_\mu\Theta^{\mu\nu}=0$ plus $\Theta^{\mu\nu}=\Theta^{\nu\mu}$ (Weinberg's version) [3] is taken as a sufficient condition, do not work for the charged metal sphere. The reason is given below.

The integral domain $V$ for the charged metal sphere is a multiply connected domain, and $V$ is surrounded by the infinite spherical surface $S_\infty(\mathbf{x})$ and the metal sphere's surface $S_{met}(\mathbf{x})$ because the electric field is equal to zero inside the metal sphere [11]. In the domain $V$, $T^{\mu\nu}$ is symmetric, and $\partial_\mu T^{\mu\nu}=0$ holds everywhere from (16) with $J_\mu=0$, but $\int T^{ij}d^3x = 0$ cannot hold. According to the Landau–Lifshitz or Weinberg's version of Laue's theorem, the time-row-element space integrals in (19) should have constituted a four-vector, but according to the sufficient and necessary condition (2) of Laue's theorem [1], they cannot constitute a four-vector because of $\int T^{ij}d^3x \neq 0$; thus resulting in the failure of both Landau–Lifshitz and Weinberg's versions of Laue's theorem.

From this analysis we find that the charged metal sphere is a typical example to show that $\partial_\mu\Theta^{\mu\nu}=0$ or $\partial_\mu\Theta^{\mu\nu}=0$ plus $\Theta^{\mu\nu}=\Theta^{\nu\mu}$ does not derive $\int T^{ij}d^3x = 0$. In other words, both the Landau–Lifshitz and Weinberg's versions of Laue's theorem break down in the case of the charged metal sphere.

As we know, the correctness of a *mathematical conjecture* cannot be legitimately affirmed by enumerating *specific* examples, no matter how many; however, it can be negated by enumerating specific examples, even only one. Thus the charged metal sphere is the right example to show that both Landau–Lifshitz and Weinberg's versions of Laue's theorem are flawed.

It is also interesting to point out that Møller's version of Laue's theorem [4], where the divergence-less plus an implicit *zero-boundary condition* is taken as a sufficient condition, does not work either in the case of the charged metal sphere, because the electric field on the boundary $S_{met}(\mathbf{x})$ is not equal to zero in terms of Coulomb's law, leading to $T^{ij}\neq 0$ on $S_{met}(\mathbf{x})$, and thus the zero-boundary condition is *not* satisfied.[10]

## 5. Application of Laue's theorem to electrostatic field confined in a finite region of space

Laue's theorem also can be directly used for analysis of an electrostatic field, which is confined in a finite electrostatic equilibrium structure, as shown in Fig. 1.

For such a finite structure, the EM stress–energy tensor $T^{\mu\nu}$ has exactly the same form as the charged metal sphere (because the fields are both distributed in *vacuum*), and satisfies Laue's pre-assumptions $\partial T^{\mu\nu}/\partial t \equiv 0$ and $T^{i4}=0$ but the sufficient and necessary condition $\int T^{ij}d^3x = 0$ cannot hold; thus resulting in the same conclusion: the total (Abraham=Minkowski) EM momentum and energy, namely, $\int(\mathbf{D}\times\mathbf{B})d^3x$ and $(1/c)\int W_{em}d^3x$, cannot constitute a four-vector although $T^{\mu\nu}$ is symmetric and $\partial_\mu T^{\mu\nu}=0$ holds (source-free). This conclusion is consistent with the specific calculations for an ideal capacitor with finite dimensions [12].[11] However according to Landau–Lifshitz or Weinberg's version of Laue's theorem, $\int(\mathbf{D}\times\mathbf{B})d^3x$ and $(1/c)\int W_{em}d^3x$ should constitute a four-vector. Thus this finite electrostatic equilibrium structure is an alternative typical example to negate both Landau–Lifshitz and Weinberg's versions of Laue's theorem.

It should be pointed out that this electrostatic equilibrium structure does not satisfy the *zero-boundary condition* required by Møller's version of Laue's theorem although it is *a finite system*.

## 6. Relation between the divergence property and the sufficient and necessary condition

In this section, we will address the relation between the divergence property of a tensor and the sufficient and necessary condition (2) in Laue's theorem that is shown in Sect. 2.

From differential rules, we have $\partial_k(\Theta^{k\nu}X^\mu)=(\partial_k\Theta^{k\nu})X^\mu+\Theta^{\mu\nu}$ and $\partial_k(\Theta^{\mu k}X^\nu)=(\partial_k\Theta^{\mu k})X^\nu+\Theta^{\mu\nu}$, where $k, \mu, \nu=1, 2, 3,$ and 4. With $\partial\Theta^{\mu\nu}/\partial t \equiv 0 \Rightarrow \partial\Theta^{\mu\nu}/\partial X^4 \equiv 0$ taken into account, from the three-dimensional Gauss divergence law we have

$$\oint_S dS_l(\Theta^{l\nu}X^i) = \int_V [(\partial_l \Theta^{l\nu})X^i]d^3x + \int_V \Theta^{i\nu}d^3x \quad (20)$$

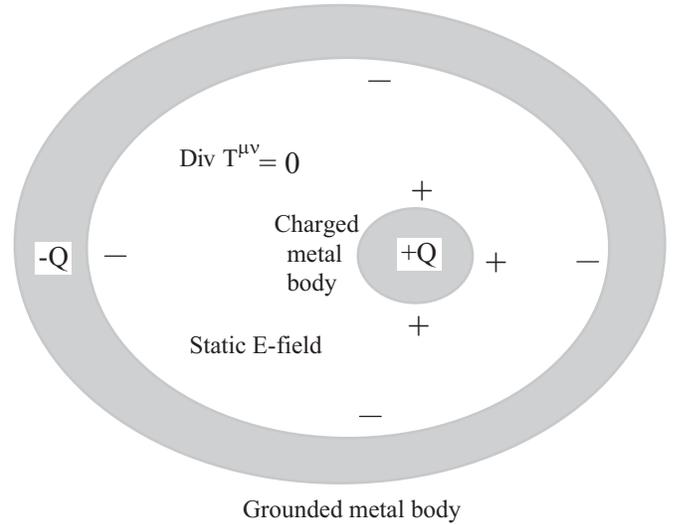

**Fig. 1.** An electrostatic field, which is confined in a finite region of space between two closed charged perfect metal bodies. According to von Laue's theorem, $\int(\mathbf{D}\times\mathbf{B})d^3x$ and $(1/c)\int W_{em}d^3x$ cannot constitute a four-vector although $T^{\mu\nu}$ is symmetric and $\partial_\mu T^{\mu\nu}=0$ holds everywhere in the field-confined region. Note that according to the EM boundary condition for perfect conductors, the electric field on the boundaries of the field-distribution region (namely, the inner surfaces of the metal body structure) is perpendicular to the boundaries [11], and it is *not* equal to zero because the bodies are charged; thus leading to $T^{\mu\nu}\neq 0$ on the boundaries; in other words, this system does not satisfy the *zero-boundary condition* in Møller's version of Laue's theorem.

---

[10]In Møller's version of Laue's theorem, an implicit zero-boundary condition is imposed, and it is combined with the divergence-less to reach the needed result that the space integrals of time-column elements are not dependent on time (corresponding to the physical implication that the total momentum and energy of an isolated system are conserved), as shown in Eq. (24) on p. 167 of ref. 4, while in the original Laue's theorem [1], the pre-assumption $\partial\Theta^{\mu\nu}/\partial t \equiv 0$ is used to reach the above needed result. Obviously, $\partial\Theta^{\mu\nu}/\partial t \equiv 0$ does not necessarily mean a "zero-boundary condition".

[11]See: Sect. "XI. Does the field momentum–energy constitute a four vector?" of ref. 12. Just like the charged metal sphere, the stress–energy tensor for an ideal planar-plate capacitor in vacuum is also *symmetric* and *divergence-less*. However, as indicated by Mansuripur and Zakharian, in an ideal planar-plate capacitor the total "field's momentum–energy is seen to behave in a way that is *not* expected from a four vector".

($i, l = 1, 2, 3$ for left-divergence and row-four-vector)

$$\oint_S dS_l(\Theta^{\mu l}X^j) = \int_V [(\partial_l \Theta^{\mu l})X^j]d^3x + \int_V \Theta^{\mu j}d^3x \quad (21)$$

($j, l = 1, 2, 3$ for right-divergence and column-four-vector)

where (20) can be used to analyze the relation between the *left*-divergence-less $\partial_\mu \Theta^{\mu\nu}(\mathbf{x}) = 0$ and the sufficient and necessary condition $\int_V \Theta^{ij}(\mathbf{x})d^3x = 0$ for the *row*-four-vector case, while (21) can be used to analyze the relation between the *right*-divergence-less $\partial_\nu \Theta^{\mu\nu}(\mathbf{x}) = 0$ and $\int_V \Theta^{ij}(\mathbf{x})d^3x = 0$ for the *column*-four-vector case. If $\Theta^{\mu\nu}(\mathbf{x})$ is symmetric ($\Theta^{\mu\nu} = \Theta^{\nu\mu}$), (20) and (21) are the same.

Without loss of generality, let us only restrict our discussions to (20) for the row-four-vector case. Note: the pre-assumption $\Theta^{i4}(\mathbf{x}) = 0$ plus (2) derives $\int_V \Theta^{i\nu}d^3x = 0$ in (20).

Generally speaking, whether $\oint_S dS_l(\Theta^{l\nu}X^i) = 0$ in (20) can hold depends on the property of specific $\Theta^{\mu\nu}(\mathbf{x})$, and we cannot affirm $\partial_\mu \Theta^{\mu\nu} = 0$ only from $\int_V \Theta^{i\nu}d^3x = 0$;[12] inversely we cannot affirm $\int_V \Theta^{i\nu}d^3x = 0$ either only from $\partial_\mu \Theta^{\mu\nu} = 0 \Rightarrow \partial_l \Theta^{l\nu} = 0$. Therefore, $\partial_\mu \Theta^{\mu\nu}(\mathbf{x}) = 0$ (divergence-less) is neither a necessary condition nor a sufficient condition of Laue's theorem.

However, the divergence-less property plus an additional boundary condition can be a sufficient condition, which is shown below.

Suppose that the boundary condition $\oint_S dS_l(\Theta^{l\nu}X^i) = 0$ holds, which is true if $\Theta^{\mu\nu}(\mathbf{x}) = 0$ is imposed on the boundary. Thus from (20), with $\partial \Theta^{\mu\nu}/\partial t \equiv 0$ and $\partial_\mu \Theta^{\mu\nu} = 0 \Rightarrow \partial_l \Theta^{l\nu} = 0$ taken into account we have $\int_V \Theta^{i\nu}d^3x = 0$, namely, $\int_V \Theta^{ij}d^3x = 0$ plus $\int_V \Theta^{i4}d^3x = 0$. $\int_V \Theta^{i4}d^3x = 0$ is automatically satisfied because of the pre-assumption $\Theta^{i4} = 0$, and $\int_V \Theta^{ij}d^3x = 0$ is the sufficient and necessary condition. Thus $\partial_\mu \Theta^{\mu\nu} = 0$ plus $\oint_S dS_l(\Theta^{l\nu}X^i) = 0$ is a *sufficient* condition of Laue's theorem. However, according to (20) we see that we cannot have both $\partial_\mu \Theta^{\mu\nu} = 0$ and $\oint_S dS_l(\Theta^{l\nu}X^i) = 0$ holding only from $\int_V \Theta^{i\nu}d^3x = 0$. In other words, $\partial_\mu \Theta^{\mu\nu} = 0$ plus $\oint_S dS_l(\Theta^{l\nu}X^i) = 0$ is not a *necessary* condition.

For the charged metal sphere, $\partial T^{\mu\nu}/\partial t \equiv 0$, $T^{i4} = 0$, and $\partial_\mu T^{\mu\nu} = 0$ are satisfied, but $\oint_S dS_l(T^{l\nu}X^i) = 0$ cannot hold for all $i$ and $\nu$ because $T^{\mu\nu} = 0$ cannot hold on the metal sphere's surface $S_{\text{met}}(\mathbf{x})$. Thus $T^{\mu\nu}$ does not satisfy the *sufficient* condition, $\partial_\mu T^{\mu\nu} = 0$ *plus* $\oint_S dS_l(T^{l\nu}X^i) = 0$. However, we cannot conclude that the total momentum and energy cannot constitute a four-vector, because $\partial_\mu T^{\mu\nu} = 0$ plus $\oint_S dS_l(T^{l\nu}X^i) = 0$ is not a *necessary* condition. (Note: A sufficient condition can be used to affirm the theorem's conclusion while it cannot be used to negate the conclusion, as stated in Sect. 1.)

## 7. Generalized von Laue's theorem for a Lorentz four-vector

Because $\int_V \Theta^{ij}(\mathbf{x})d^3x = 0$ and $\Theta^{i4}(\mathbf{x}) = 0$ derive $\int_V \Theta^{i\nu}(\mathbf{x})d^3x = 0$ ($i, j = 1, 2, 3; \nu = 1, 2, 3, 4$), the latter is a looser condition. It is easy to show that the sufficient and necessary condition (2) can be replaced by $\int_V \Theta^{i\nu}(\mathbf{x})d^3x = 0$, to obtain a generalized von Laue's theorem, as follows.

**Generalized von Laue's theorem.** Assume that $\Theta^{\mu\nu}(\mathbf{x})$ is a Lorentz four-tensor given in the laboratory frame $XYZ$ ($\mu, \nu = 1, 2, 3$, and 4, with the index 4 corresponding to time component), and $\Theta^{\mu\nu}$ is independent of time ($\partial \Theta^{\mu\nu}/\partial t \equiv 0$). The generalized von Laue's theorem states: *The time-row-element space integrals*

$$P^\nu = \int_V \Theta^{4\nu}d^3x \quad (22)$$

*constitute a Lorentz four-vector* if and only if

$$\int_V \Theta^{i\nu}d^3x = 0 \quad (23)$$

*holds for all* $i = 1, 2, 3$, *and* $\nu = 1, 2, 3, 4$.

*Proof:* From (6) before the pre-assumption $\Theta^{i4} = 0$ is used, we have

$$P'^\nu = \frac{\partial X'^\nu}{\partial X^\lambda}P^\lambda + \frac{\partial X'^\nu}{\partial X^\lambda}\frac{\partial X'^4}{\partial X^i}\gamma^{-1}\int_V \Theta^{i\lambda}(\mathbf{x})d^3x \quad (24)$$

The sufficient and necessary condition for the validity of $P'^\nu = (\partial X'^\nu/\partial X^\lambda)P^\lambda$ is the holding of

$$\gamma^{-1}\frac{\partial X'^\nu}{\partial X^\lambda}\left[\int_V \Theta^{i\lambda}(\mathbf{x})d^3x\right]\frac{\partial X'^4}{\partial X^i} = 0 \quad (25)$$

or

$$\begin{pmatrix} 1+\xi\beta_x^2 & \xi\beta_x\beta_y & \xi\beta_x\beta_z & -\gamma\beta_x \\ \xi\beta_y\beta_x & 1+\xi\beta_y^2 & \xi\beta_y\beta_z & -\gamma\beta_y \\ \xi\beta_z\beta_x & \xi\beta_z\beta_y & 1+\xi\beta_z^2 & -\gamma\beta_z \\ -\gamma\beta_x & -\gamma\beta_y & -\gamma\beta_z & \gamma \end{pmatrix}$$
$$\times \begin{pmatrix} a_{11} & a_{12} & a_{13} \\ a_{21} & a_{22} & a_{23} \\ a_{31} & a_{32} & a_{33} \\ a_{41} & a_{42} & a_{43} \end{pmatrix}\begin{pmatrix} -\gamma\beta_x \\ -\gamma\beta_y \\ -\gamma\beta_z \end{pmatrix} = \begin{pmatrix} 0 \\ 0 \\ 0 \\ 0 \end{pmatrix} \quad (26)$$

where $a_{\lambda i} = \int_V \Theta^{i\lambda}(\mathbf{x})d^3x$, with $\lambda = 1, 2, 3, 4$ and $i = 1, 2, 3$. The sufficiency of (23) is apparent. The necessity is based on the fact: $\boldsymbol{\beta}c$ is arbitrary, and $a_{\lambda i} = 0$ must hold for all $\lambda$ and $i$, because ($\beta_x \neq 0, \beta_y = 0, \beta_z = 0$) $\Rightarrow a_{\lambda 1} = 0$, ($\beta_x = 0, \beta_y \neq 0, \beta_z = 0$) $\Rightarrow a_{\lambda 2} = 0$, and ($\beta_x = 0, \beta_y = 0, \beta_z \neq 0$) $\Rightarrow a_{\lambda 3} = 0$.

From (20) we see that $\partial_\mu \Theta^{\mu\nu} = 0$ (divergence-less) plus $\oint_S dS_l(\Theta^{l\nu}X^i) = 0$ (boundary condition) derives $\int_V \Theta^{i\nu}d^3x = 0$, which is the sufficient and necessary condition given by (23). In other words, the divergence-less plus an additional boundary condition is also a sufficient condition for the generalized Laue's theorem. But of course, it is not a necessary condition.

Like the classical Laue's theorem presented in Sect. 2, we also can designate the time-*column*-element space integrals as a four-vector, for which the statement is modified into: Assume that $\Theta^{\mu\nu}(\mathbf{x})$ is a Lorentz four-tensor given in the laboratory frame $XYZ$ ($\mu, \nu = 1, 2, 3$, and 4, with the index 4 corresponding to time component), and $\Theta^{\mu\nu}$ is independent of time ($\partial \Theta^{\mu\nu}/\partial t \equiv 0$). The generalized von Laue's theorem states: *The time-column-element space integrals* $\Pi^\mu = \int_V \Theta^{\mu 4}d^3x$ *constitute a Lorentz four-vector* if and only if $\int_V \Theta^{\mu j}d^3x = 0$ *holds for all* $j = 1, 2, 3$, *and* $\mu = 1, 2, 3, 4$.

---

[12]For simplicity and clarity, let us take a three-dimensional-vector as an example to illustrate that $\int_V \Theta^{ij}(\mathbf{x})d^3x = 0$ does not derive $\partial_i \Theta^{ij}(\mathbf{x}) = 0$. For example, suppose a vector is given by $\mathbf{A} = \mathbf{x}$ in the cube $|x, y, z| \leq 1$ and $\mathbf{A} = 0$ outside of the cube. Thus we have the space integral $\int \mathbf{A}d^3x = 0$ over the whole space, but $\nabla \cdot \mathbf{A} = 3 \neq 0$ in the cube, which means $\int \mathbf{A}d^3x = 0$ does not derive $\nabla \cdot \mathbf{A} = 0$. Recall that the charged metal sphere example has shown that $\partial_\mu \Theta^{\mu\nu}(\mathbf{x}) = 0$ does not derive $\int_V \Theta^{ij}(\mathbf{x})d^3x = 0$. Therefore, we can conclude that $\partial_\mu \Theta^{\mu\nu}(\mathbf{x}) = 0$ is neither a sufficient condition nor a necessary condition for the Laue's theorem.

## 8. Derivative von Laue's theorem for a Lorentz invariant

Now let us introduce the derivative von Laue's theorem for a Lorentz invariant. The way to prove this is the same as that for a Lorentz four-vector.

**Derivative von Laue's theorem.** Assume that $\Lambda^\mu(\mathbf{x}) = (\mathbf{\Lambda}, \Lambda^4)$ is a Lorentz four-vector given in the laboratory frame $XYZ$ ($\mu = 1, 2, 3$, and 4, with the index 4 corresponding to time element), and $\Lambda^\mu$ is independent of time ($\partial \Lambda^\mu/\partial t \equiv 0$). von Laue's theorem states: *The time-element space integral*

$$\Psi = \int_V \Lambda^4(\mathbf{x}) d^3 x \tag{27}$$

*is a Lorentz invariant* if and only if

$$\int_V \Lambda^i(\mathbf{x}) d^3 x = 0 \quad (i = 1, 2, 3) \quad \text{or} \quad \int_V \mathbf{\Lambda}(\mathbf{x}) d^3 x = 0 \tag{28}$$

*holds.*

*Proof*: Suppose that $X'Y'Z'$ is an inertial frame moving at $\boldsymbol{\beta} c$ with respect to the laboratory frame $XYZ$. According to the Lorentz transformation of $\Lambda'^\mu$ from $\Lambda^\mu$, we have

$$\begin{aligned}\Psi' &= \int_{V':\, t'=\text{const}} \Lambda'^4(\mathbf{x}', ct') d^3 x' = \frac{\partial X'^4}{\partial X^\lambda} \int_{V':\, t'=\text{const}} \Lambda^\lambda(\mathbf{x} = \mathbf{x}(\mathbf{x}', ct')) d^3 x' = \frac{\partial X'^4}{\partial X^\lambda} \frac{1}{\gamma} \int_V \Lambda^\lambda(\mathbf{x}) d^3 x \\ &= \frac{\partial X'^4}{\partial X^4} \frac{1}{\gamma} \int_V \Lambda^4(\mathbf{x}) d^3 x + \frac{\partial X'^4}{\partial X^i} \frac{1}{\gamma} \int_V \Lambda^i(\mathbf{x}) d^3 x = \int_V \Lambda^4(\mathbf{x}) d^3 x - \boldsymbol{\beta} \cdot \int_V \mathbf{\Lambda}(\mathbf{x}) d^3 x = \Psi - \boldsymbol{\beta} \cdot \int_V \mathbf{\Lambda}(\mathbf{x}) d^3 x \end{aligned} \tag{29}$$

From (29) it is apparent that (28) is a sufficient condition for $\Psi' = \Psi$, and it is also a necessary condition because $\boldsymbol{\beta}$ is arbitrary.

As an application of this kind of Laue's theorem, the Lorentz invariance of total charge is shown below.

Suppose that a single point charge of $q$ is fixed at $\mathbf{x} = 0$. In the point-charge rest frame, the four-current density is given by $J^\mu(\mathbf{x}) = (\mathbf{J}, J^4)$ with $\mathbf{J} = 0$ and $J^4 = c\rho = cq\delta(\mathbf{x})$ (where $\delta(\mathbf{x})$ is the Dirac delta function), satisfying $\partial J^\mu/\partial t \equiv 0$ and $\int \mathbf{J} d^3 x = 0$. Thus according to Laue's theorem, $\int J'^4 d^3 x' = \int J^4 d^3 x$ is a Lorentz invariant, and we have $\int c\rho' d^3 x' = \int c\rho d^3 x \Rightarrow cq' = cq$, namely, the charge $q' = \int \rho' d^3 x' = q$, observed in any inertial frame, is a Lorentz invariant.

This conclusion can be easily generalized for general cases. Observed in any specific inertial frame, the total charge density is a sum of the densities of all point charges, given by $\rho' = \sum \rho'_i$. For all individual point charges, $q'_i = \int \rho'_i d^3 x' = q_i$ holds, and thus the total charge

$$Q' = \int \rho' d^3 x' = \sum_i \int \rho'_i d^3 x' = \sum_i q'_i = \sum_i q_i = Q \tag{30}$$

is a Lorentz invariant, although each of the point charges may have its own charge rest frame.

According to this analysis we can see that the Lorentz invariance of total charge results from the two facts:

1. The current density $J^\mu(\mathbf{x}) = (\mathbf{J}, J^4)$ is a Lorentz four-vector, which is required by the invariance of Maxwell equation $\partial_\mu G^{\mu\nu} = J^\nu$.[13]
2. The moving velocity of any point charge is less than light speed $c$ so that there is a point-charge rest frame where $\partial J^\mu/\partial t \equiv 0$ and $(\mathbf{J} = 0 \Rightarrow) \int \mathbf{J} d^3 x = 0$ hold.

Nevertheless the total charge is usually taken as an experimental invariant, as indicated by Jackson [5, p. 555].

## 9. Summary

In this paper, von Laue's theorem, as well as its generalized form, is strictly proved for its sufficient and necessary condition (SNC). It is shown that the divergence-less property of a four-tensor itself is neither a sufficient nor a necessary condition, while the divergence-less *plus* an additional boundary condition only can be a sufficient condition, which *cannot* be used to judge the Lorentz property of the total EM momentum and energy for a charged metal sphere in free space. As an application, the SNC version of Laue's theorem is used to analyze the infinitely extended electrostatic field produced by a charged metal sphere, and the static field confined in a finite electrostatic equilibrium structure. A derivative von Laue's theorem is introduced, and it is employed to show the Lorentz invariance of total charge.

It should be noted that, in the original work [1], Laue only provided a proof for its sufficient condition. In principle, this original Laue's theorem cannot be used to analyze the Lorentz property of EM momentum and energy generated by a charged metal sphere or confined in a finite electrostatic equilibrium structure, because the EM stress–energy tensor does not meet Laue's sufficient condition. Thus the SNC version of Laue's theorem provided in the paper overcomes the difficulty of the original Laue's theorem.

It is found in the paper that the Landau–Lifshitz version of Laue's theorem (where the divergence-less of a four-tensor is taken as a sufficient condition) and Weinberg's version of Laue's theorem (where the divergence-less plus a symmetry is taken as a sufficient condition) are both flawed, although they are widely accepted as well-established basic results of tensor calculus [2, 3]. That is because the two versions of Laue's theorem are directly negated by the specific examples of the charged metal sphere and the finite electrostatic equilibrium structure, for which the EM stress–energy tensor is both *symmetric* and *divergence-less*, but the space integrals of the time-row (column) elements of the tensor *cannot* constitute a Lorentz four-vector, as shown in Sects. 4 and 5.

It is also found that, the Møller's version of Laue's theorem [4], where the divergence-less plus a *zero-boundary condition* is taken as a sufficient condition, has a very limited application. For example, as shown in Sect. 4, Møller's version of Laue's theorem cannot be used to analyze the charged metal sphere, and as shown in Sect. 5, it cannot even be used to judge the Lorentz property of EM momentum and energy confined in an electrostatic equilibrium structure — a finite system, the argument on which the imposed "zero-boundary condition" is based.

---

[13] As shown in Sect. 11. 9 of the textbook by Jackson [5], Maxwell equations $[\nabla \times \mathbf{H} - \partial(c\mathbf{D})/\partial(ct), \nabla \cdot (c\mathbf{D})] = (\mathbf{J}, c\rho)$ and $[\nabla \times \mathbf{E} - \partial(-c\mathbf{B})/\partial(ct), \nabla \cdot (-c\mathbf{B})] = (\mathbf{0}, 0)$ can be written as $\partial_\mu G^{\mu\nu} = J^\nu$ and $\partial_\mu \mathcal{F}^{\mu\nu} = 0$, where $\mathcal{F}^{\mu\nu}$ is the dual field-strength tensor. $G^{\mu\nu}$ and $\mathcal{F}^{\mu\nu}$ are assumed to be four-tensor Lorentz covariant to keep Maxwell equations invariant in form in all inertial frames, and thus $(\mathbf{J}, c\rho)$ must be a four-vector.

# Materials to help reading

For the convenience of readers, illustrations are given below to the footnotes in my manuscript.

---------------------------------------------------------------------------------------------

**Footnote-1:**

[1] In ref. 2, the book by Landau and Lifshitz, the *divergence-less* of a tensor is taken as a sufficient condition, as shown in Eq. (32.6) on p. 83 and Eq. (32.11) on p. 84. The *symmetry* of the tensor is claimed to be required by "the law of conservation of angular momentum" by repeating use of their version of Laue's theorem; see Eq. (32.10) on p. 84. As shown in Sect. 4 of the present paper, however, the divergence-less is *never* a sufficient condition; thus the correctness of the requirement of the symmetry is also questionable.

The comments in footnote-1 are based on the text in Ref. [2] (L. D. Landau and E. M. Lifshitz, The Classical Theory of Fields, Butterworth-Heinemann, New York, 1975), copied below:

$$\frac{\partial T_i^k}{\partial x^k} = 0. \tag{32.4}$$

We note that if there is not one but several quantities $q^{(l)}$, then in place of (32.3) we must write

$$T_i^k = \sum_l q_{,i}^{(l)} \frac{\partial \Lambda}{\partial q_{,k}^{(l)}} - \delta_i^k \Lambda. \tag{32.5}$$

But in § 29 we saw that an equation of the form $\partial A^k/\partial x^k = 0$, i.e. the vanishing of the four-divergence of a vector, is equivalent to the statement that the integral $\int A^k dS_k$ of the vector over a hypersurface which contains all of three-dimensional space is conserved. It is clear that an analogous result holds for the divergence of a tensor; the equation (32.4) asserts that the vector $P^i = \text{const} \int T^{ik} dS_k$ is conserved.

This vector must be identified with the four-vector of momentum of the system. We choose the constant factor in front of the integral so that, in accord with our previous definition, the time component $P^0$ is equal to the energy of the system multiplied by $1/c$. To do this we note that

$$P^0 = \text{const} \int T^{0k} dS_k = \text{const} \int T^{00} dV$$

if the integration is extended over the hyperplane $x^0 = $ const. On the other hand, according to (32.3),

$$T^{00} = \dot{q}\frac{\partial \Lambda}{\partial \dot{q}} - \Lambda. \quad \left(\dot{q} \equiv \frac{\partial q}{\partial t}\right)$$

Comparing with the usual formulas relating the energy and the Lagrangian, we see that this quantity must be considered as the energy density of the system, and therefore $\int T^{00} dV$ is the total energy of the system. Thus we must set const = $1/c$, and we get finally for the four-momentum of the system the expression

<span style="color:red">This defintion itself is a 4-vector.</span>
$$P^i = \frac{1}{c}\int T^{ik} dS_k. \tag{32.6}$$

The tensor $T^{ik}$ is called the *energy-momentum tensor* of the system.

As we mentioned above, if we carry out the integration in (32.6) over the hyperplane $x^0 = $ const., then $P^i$ takes on the form

<span style="color:red">However, there is no specific proof about how Eq. (32.6) can become Eq. (32.11).</span>
$$P^i = \frac{1}{c}\int T^{i0} dV, \tag{32.11}$$

where the integration extends over the whole (three-dimensional) space. The space components of $P^i$ form the three-dimensional momentum vector of the system and the time component is its energy multiplied by $1/c$. Thus the vector with components

To define the tensor $T^{ik}$ uniquely we can use the requirement that the four-tensor of angular momentum (see § 14) of the system be expressed in terms of the four-momentum by

$$M^{ik} = \int (x^i dP^k - x^k dP^i) = \frac{1}{c}\int (x^i T^{kl} - x^k T^{il}) dS_l, \quad (32.8)$$

that is its "density" is expressed in terms of the "density" of momentum by the usual formula.

It is easy to determine what conditions the energy-momentum tensor must satisfy in order that this be valid. We note that the law of conservation of angular momentum can be expressed, as we already know, by setting equal to zero the divergence of the expression under the integral sign in $M^{ik}$. Thus

$$\frac{\partial}{\partial x^l}(x^i T^{kl} - x^k T^{il}) = 0. \quad (32.9)$$

Noting that $\partial x^i/\partial x^l = \delta_l^i$ and that $\partial T^{kl}/\partial x^l = 0$, we find from this

$$\delta_l^i T^{kl} - \delta_l^k T^{il} = T^{ki} - T^{ik} = 0$$

or

$$T^{ik} = T^{ki}, \quad (32.10)$$

that is, the energy-momentum tensor must be symmetric.

[Annotation: By repeating use of their own version of Laue's theorem, they proved the energy-momentum tensor must be symmetric.]

From their own version of Laue's theorem, Landau and Lifshitz claim that the law of conservation of angular momentum can be expressed by setting the divergence of $(x^i T^{kl} - x^k T^{il})$ in Eq. (32.8) to be zero. In other words, according to their own version of Laue's theorem, the divergence-less of $(x^i T^{kl} - x^k T^{il})$ means that the angular momentum, the space integrals of time-column elements of $(x^i T^{kl} - x^k T^{il})$, is constant (conserved). Thus they proved "the energy-momentum tensor must be symmetric", namely Eq. (32.10). Note that the space integrals of time-column elements of $(x^i T^{kl} - x^k T^{il})$ are the time-column elements of the four-tensor of angular momentum $M^{ik}$.

------------------------------------------------------------------------------------------------

**Footnote-2:**

[2]On p. 46 of ref. 3, Weinberg argues that it "can be shown" that the divergence-less and symmetry is a sufficient condition.

The comments in footnote-2 are based on the text in Ref. [3] (S. Weinberg, Gravitation and Cosmology: Principles and Applications of the General Theory of Relativity, John Wiley & Sons, New York, 1972), copied below:

This is again a symmetric tensor, and is now conserved

$$\partial_\alpha T^{\alpha\beta} = 0 \quad (2.8.13)$$

46    2 Special Relativity

We can continue to add more and more terms to $T^{\alpha\beta}$ to account for other fields and keep $T^{\alpha\beta}$ conserved. A systematic method for constructing these terms is presented in Chapter 12.

Just as the integral of the charge density $J^0$ is the total charge, the integral of the density $T^{\alpha 0}$ of $p^\alpha$ is the total $p^\alpha$:

$$p^\alpha_{total} = \int d^3x\, T^{\alpha 0}(x,t) \quad (2.8.14)$$

That this is a constant four-vector can be shown in the same way that we showed in Section 6 that the total charge (2.6.7) is a constant scalar.

---

**Footnote-3:**

[3]On pp. 166–169 of ref. 4, Møller provided a proof that the divergence-less property plus an implicit "zero-boundary condition" is a sufficient condition. This zero-boundary condition, combined with the divergence-less, insures that the space integrals of the time-column elements "are constant in time", as shown in Eq. (24) on p. 167. The argument of the zero-boundary condition is "the system considered is finite"; however, in practice, a finite system does not necessarily mean a zero-boundary condition, of which a typical example is given in Fig. 1 of the present paper. Thus the use of Møller's version of Laue's theorem is very limited; for example, it cannot be used to judge the Lorentz property of EM momentum and energy of the charged metal sphere; see Sect. 4 of the present paper.

The comments in footnote-3 are based on the text in Ref. [4] (C. Møller, The Theory of Relativity, Oxford University Press, London, 1955), copied below:

fields by analogy with (V. 105). For the total system of matter and fields we then get, in the same way as in § 61, the laws of conservation of energy and momentum in the form

$$\frac{\partial T_{ik}}{\partial x_k} = 0, \qquad (1)$$

where $T_{ik}$ is the total energy-momentum tensor of the closed system.

Let us now assume that the system considered is finite so that all components $T_{ik}$ are zero outside a certain region in physical space. If we multiply (1) by $dx_1 dx_2 dx_3$ and integrate over the whole physical space

VI, § 63   MECHANICS OF ELASTIC CONTINUA   167

for constant $x_4$, the first three terms in (1), which are partial derivatives of $T_{ik}$ with respect to the space coordinates $x_\kappa$, will give zero. Hence we get

$$\frac{d}{dx_4} \int T_{i4} dV = 0,$$

To get this, an implicit zero-boundary condition is used because "the system considered is finite".

which shows that the four quantities

$$G_i = \int g_i dV = \left(\mathbf{G}, \frac{i}{c}H\right) \qquad (24)$$

are constant in time G and H represent the total linear momentum and the total energy of the system, respectively. Another consequence of (1) is that the quantities $G_i$ transform like the components of a four-vector, the four-momentum vector. This is seen in the following way.

---

## Footnote-4:

[4] On p. 756 of ref. 5, Jackson presented Landau–Lifshitz version of Laue's theorem [2], where the divergence-less described by Eq. (16.39) is taken as a sufficient condition, and it is thought to be equivalent to the original Laue's sufficient condition Eq. (16.40). In fact, Eq. (16.39) does not necessarily mean Eq. (16.40), as shown in Sect. 4 of the present paper.

The comments in footnote-4 are based on the text in Ref. [5] (J. D. Jackson, Classical Electrodynamics, John Wiley & Sons, NJ, 1999), copied below:

As Poincaré observed in 1905–1906,* a deficiency of the purely electromagnetic classical models is their lack of stability. Nonelectromagnetic forces are necessary to hold the electric charge in place. Poincaré therefore proposed such forces, described by a stress tensor $P^{\alpha\beta}$ to be added to the electromagnetic $\Theta^{\alpha\beta}$ to give a total stress tensor $S^{\alpha\beta}$,

$$S^{\alpha\beta} = \Theta^{\alpha\beta} + P^{\alpha\beta}$$

The particle's total 4-momentum is then defined to be

$$cP^{\alpha} = \int S^{\alpha 0}\, d^3x \qquad (16.38)$$

where the integral is over all 3-space at a fixed time. The right-hand side of (16.38) transforms as a 4-vector provided

This is Landau-Lifshitz sufficient condition.
$$\partial_{\alpha} S^{\alpha\beta} = 0 \qquad (16.39)$$

or equivalently, provided

This is Laue's original sufficient condition.
$$\int S^{(0)ij}\, d^3x^{(0)} = 0 \qquad (16.40)$$

with $i, j = 1, 2, 3$, and the superscript (0) denoting the rest frame ($\mathbf{P} = 0$). Condition (16.40) is just the statement that the total self-stress (in the three-dimensional sense) must vanish—the condition for mechanical stability.

---

## Footnote-5:

[5] In ref. 7, the divergence-less is claimed as a sufficient condition in Eq. (19), while it is taken as the necessary condition in Eq. (51).

The comments in footnote-5 are based on the text in Ref. [7] (D. J. Griffiths, Am. J. Phys. 80 (2012) 7), copied below:

where **v** is the velocity of the charge and **a** is its acceleration. [The uniqueness of these expressions [Eqs. (7)–(10)] is open to some question (Ref. 23), but I shall take them as definitions.]

Several conservation laws follow from Maxwell's equations. Local conservation of charge is expressed by the continuity equation,

$$\frac{\partial \rho}{\partial t} + \nabla \cdot \mathbf{J} = 0. \qquad (13)$$

not have their electromagnetic form [Eqs. (7)–(10)]. If the stress-energy tensor is divergenceless:

$$\partial_\mu \Theta^{\mu\nu} = 0 \qquad \text{sufficient condition} \qquad (19)$$

then

$$p^\mu \equiv \int \Theta^{0\mu}\, d^3\mathbf{r} \qquad (20)$$

transforms as a four-vector [this is sometimes called "von Laue's theorem" (Refs. 24, 26, and 7)], and the total energy and momentum ($E = cp^0$ and **p**) are conserved. If the stress   a conserved four-vector. And (as always) the *complete*

More formally, the problem is that the electromagnetic stress-energy tensor is not divergenceless in the presence of charge and current [Eq. (29)], and as a result the integral

$$p_{em}^\mu = \int T^{0\mu} d^3\mathbf{r} \quad (51)$$

The divergenceless is taken as a necessary condition here.

does not constitute a four-vector. It is only the *complete* stress-energy tensor (which in this instance would include a contribution from the Poincaré stress) that is divergenceless, and *its* integral does yield a genuine (and conserved) four-vector (Eq. (20)).

---

## Footnote-6:

[6] In ref. 10, as shown in Eq. (5), the symmetry and divergence-less of a four-tensor is implicitly taken as a sufficient condition for the space integrals of the time-column elements to constitute a Lorentz four-vector; namely, the Weinberg's version of Laue's theorem.

The comments in footnote-6 are based on the text in Ref. [10] (T. Ramos, G. F. Rubilar, Y. N. Obukhov, Phys. Lett. A 375 (2011) 1703; http://arxiv.org/abs/1103.1654 ), copied below:

T. Ramos et al. / Physics Letters A 375 (2011) 1703–1709

The total tensor is symmetric and satisfies the following energy–momentum balance equation,

$$\partial_\nu T_\mu{}^\nu - F_{\mu\nu} J^\nu_{ext} = 0, \quad (4)$$

where the 4-vector $J^\nu_{ext}$ describes the external charge and current densities which do not belong to the dielectric fluid. If $J^\nu_{ext} = 0$ energy–momentum tensor of the complete system is conserved and we have a closed system.

If we choose a volume $V'$ big enough so that it encloses the pulse and the slab until the pulse leaves the slab from the other side, then we can integrate the conservation equation and obtain that the total 4-momentum $\mathcal{P}_\mu := (E, -\mathbf{p})$ of the whole system, defined as

$$\mathcal{P}_\mu := \int_{V'} T_\mu{}^0 dV, \quad (5)$$

is a conserved, i.e. time independent, quantity. We will use this

Copied paragraphs to show the implicit assumption made by Ramos, Rubilar, and Obukhov: If a Lorentz 4-tensor is symmetric and divergence-less, then the integrals of time-column elements constitute a Lorentz 4-vector, namely Weinberg's version of Laue's theorem.

Note: $\partial_\nu T_\mu{}^\nu - F_{\mu\nu} J^\nu_{ext} = 0$ and $J^\nu_{ext} = 0$ $\Rightarrow \partial_\nu T_\mu{}^\nu = 0$ (divergence-less).